\documentclass{jps-cp}
\usepackage{graphicx}
\usepackage{here}

\title{Optical Conductivity Study of $f$ Electron States in 
YbCu$_2$Ge$_2$ at High Pressures to 20~GPa}

\author{Hidekazu \textsc{Okamura}$^1$, 
Makoto \textsc{Nagata}$^2$, 
Atsushi \textsc{Tsubouchi}$^1$, 
Yoshichika \textsc{$\bar{\rm O}$nuki}$^3$, 
Yuka \textsc{Ikemoto}$^4$, and 
Taro \textsc{Moriwaki}$^4$
}

\inst{$^1$ Graduate School of Advanced Technology and Science, 
Tokushima University, Tokushima 770-8506, Japan \\
$^2$ Graduate School of Science, Kobe University, Kobe 657-8501, 
Japan \\
$^3$ Faculty of Science, University of the Ryukyus, 
Okinawa 903-0213, Japan \\
$^4$ Japan Synchrotron Radiation Research Center, Hyogo 679-5198, 
Japan
}

\email{ho@tokushima-u.ac.jp}

\recdate{September 10, 2019}

\abst{Optical conductivity [$\sigma(\omega)$] of YbCu$_2$Ge$_2$ 
has been measured at external pressures ($P$) to 20~GPa, 
to study the $P$ evolution of $f$ electron hybridized states.   
At $P$=0, $\sigma(\omega)$ shows a marked mid-infrared 
(mIR) peak at 0.37~eV, which is due to optical 
excitations from $f^{14}$ (Yb$^{2+}$) state located 
below the Fermi level.  
With increasing $P$, the mIR peak shows significant 
shifts to lower energy, reaching 0.18~eV at $P$=20~GPa.  
This result indicates that the $f^{14}$ energy level 
increases toward the Fermi level with $P$.   Such a shift 
of the $f$ electron level with $P$ has been 
expected from theoretical considerations, but 
had never been demonstrated by spectroscopic 
experiment under high $P$.   The obtained 
results are also analyzed in terms of the $P$ 
evolution of the conduction-$f$ electron hybridization.  
}

\kword{$f$ electron hybridized state, high pressure, 
optical conductivity}

\begin{document}
\maketitle

\section{Introduction}
Duality between localized and delocalized characteristics 
exhibited by the $f$ electrons has been a central issue 
in the physics of heavy fermion (HF) compounds, which 
are typically Ce- or Yb-containing metals.  
Since the $f$ orbitals are much more localized near the 
nucleus than the conduction ($c$) electron states, the $f$ 
electrons intrinsically have localized characteristics.  
However, they may become partly delocalized 
through a hybridization with the $c$ electrons.  
The degree of localization/delocalization depends on 
the strength of the hybridization.  
Strongly localized (weakly hybridized) $f$ electrons 
lead to interesting phenomena such as effective mass 
($m^\ast$) enhancement and magnetic order at low 
temperatures ($T$).  
On the other hand, strongly delocalized (strongly 
hybridized) $f$ electrons lead to intermediate 
valence (IV) states, where the average Ce (Yb) 
valence takes a non-integer value between 3 and 4 
(2 and 3).   
The delocalized systems, in particular Yb-based 
IV compounds, are also quite interesting since their 
electronic states can be tuned by external pressure 
($P$) over a wide range from delocalized to 
localized ones.  
Since an Yb$^{3+}$ has a smaller ionic radius than 
an Yb$^{2+}$, an external $P$ generally drives the 
electronic states of an IV Yb compound toward 
Yb$^{3+}$ state.  
YbCu$_2$Ge$_2$ \cite{onuki,miyake,miyake-XAS,PES} 
is a good example of such Yb compounds: At $P$=0, 
YbCu$_2$Ge$_2$ shows a Pauli paramagnetism, with a 
specific heat coefficient ($\gamma$) of 10~mJ, 
and Yb is expected to be almost divalent \cite{onuki}.  
With increasing $P$, however, the Yb 
valence increases \cite{miyake-XAS} and $A$, 
the $T^2$ coefficient of resistivity $\rho(T)$, 
also significantly increases, indicating a large 
enhancement in $m^\ast$ \cite{miyake}.  
These results show that the $f$ electron states are 
strongly affected by $P$, and they become more 
localized at high $P$.

To study the microscopic $c$-$f$ hybridized electronic 
states in HF compounds, the optical conductivity 
[$\sigma(\omega)$] technique has played important 
roles \cite{wang}.   
Typically, $\sigma(\omega)$ of a Ce or Yb HF compound 
shows a marked mid-infrared (mIR) peak, which has been 
discussed in terms of the $c$-$f$ hybridized electronic 
states near the Fermi level ($E_{\rm F}$).    
In particular, it has been found that the mid-IR peak 
energy ($E_{\rm mIR}$) is roughly proportional to the 
$c$-$f$ hybridization energy over many Ce and Yb 
HF compounds \cite{okamura}.   
Therefore, $\sigma(\omega)$ study under high $P$ is 
expected to be a useful probe for the electronic states 
in YbCu$_2$Ge$_2$ at high $P$.  
In this work, we have measured $\sigma(\omega)$ of 
YbCu$_2$Ge$_2$ at high $P$ to 20~GPa.

\section{Experimental}
The samples of YbCu$_2$Ge$_2$ used were 
single crystals grown with self-flux method \cite{onuki}.  
The optical reflectance spectra at high $P$ were 
measured using a diamond anvil cell 
(DAC)\cite{okamura-pressure}.  
Type IIa diamond anvils with 0.8 and 0.6~mm culet 
diameter and a stainless steel gasket were used to 
seal the sample with NaCl as the pressure transmitting 
medium.   
A flat, as-grown surface of a sample was directly 
attached on the culet surface of the diamond anvil, 
and the reflectance at the sample/diamond interface, 
denoted as $R_d(\omega)$, was measured.   
A gold film was placed between the gasket and 
anvil as a reference of $R_d(\omega)$.     
Small ruby pieces were also sealed to monitor the 
pressure via its fluorescence.   
A part of the measurements was made with synchrotron 
radiation as a bright IR source\cite{JPSJ-review} at 
the beamline BL43IR of SPring-8\cite{micro1,micro2}.  
$\sigma(\omega)$ was derived from $R_d(\omega)$ using 
the Kramres-Kronig (KK) analysis\cite{dressel}.  
In the KK analysis of $R_d(\omega)$, the refractive 
index of diamond ($n_d$=2.4) was taken into account 
as previously described\cite{kk-dia}.  
Below the measured energy range, $R_d(\omega)$ was 
extrapolated with the Hagen-Rubens 
function\cite{dressel}.  
More details of the high pressure IR experiments 
can be found elsewhere\cite{okamura-pressure}.

\section{Results and Discussions}
Figure~1(a) shows $R_d(\omega)$ of YbCu$_2$Ge$_2$ 
measured with DAC at 295~K and at high $P$ to 
20~GPa.  
\begin{figure}[t]
\includegraphics[width=0.6\textwidth]{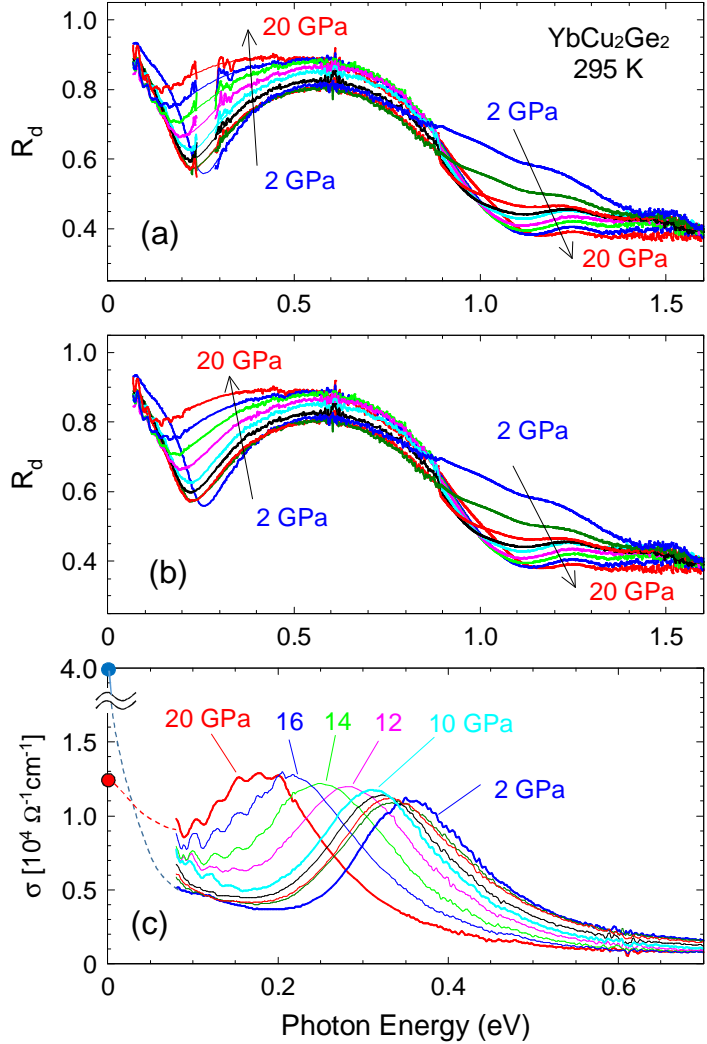}
\caption{(a)Reflectance spectra ($R_d$) of YbCu$_2$Ge$_2$ 
measured with DAC at 295~K and at $P$=2, 4, 6, 8, 10, 
12, 14, 16, and 20~GPa.   The thick solid curves show 
the original data, where a spectral range around 
0.27~eV is not indicated since this range could not 
be measured well due to strong absorption by the diamond.  
The thin solid curves show the interpolations, which 
smoothly connect the spectra at both sides of the 
diamond absorption range.  
(b) The interpolated $R_d$ spectra alone are shown.  
(c) Optical conductivity ($\sigma$) spectra of 
YbCu$_2$Ge$_2$ at high $P$ obtained by the 
Kramers-Kronig analysis of the reflectance spectra 
in (b).  The two dots on the vertical axis indicate 
the dc conductivities at 2~GPa (blue) and 20~GPa 
(red) from Ref.~\cite{miyake} and the broken curves 
are guide to the eye. 
}
\end{figure}
The thick, solid curves indicate the original 
reflectance data, but the data in the range 
0.24-0.3~eV are not indicated.  This is because 
this range could not be measured well due to 
strong absorption by the diamond.    
To complete the spectra, therefore, they were 
smoothly interpolated as indicated by the thin 
solid curves in Fig.~1(a), and the interpolated 
spectra alone are again shown in Fig.~1(b).  
It is clear that $R_d(\omega)$ of YbCu$_2$Ge$_2$ 
has significant $P$ dependences over the entire 
measured spectral range.  At 2~GPa, $R_d(\omega)$ 
has a deep minimum centered near 0.27~eV, 
then it gradually decreases 
with increasing photon energy.   In the deep 
minimum portion below 0.5~eV, with increasing $P$, 
$R_d(\omega)$ increases and the minimum becomes 
shallower and shifts to lower 
photon energy.  These $P$ evolutions occur much 
more rapidly above 10~GPa than below 10~GPa.  
At the same time, in the spectral range above 
0.9~eV, $R_d(\omega)$ decreases with $P$.  In contrast 
to the $P$ evolution below 0.5~eV discussed above, 
that above 0.9~eV is much more rapid below 10~GPa 
than above 10~GPa.

Figure~1(c) shows $\sigma(\omega)$ spectra of YbCu$_2$Ge$_2$ 
at high $P$, obtained from the $R_d(\omega)$ spectra in Fig.~1(b) 
using the KK analysis.  
The main feature in $\sigma(\omega)$ is a pronounced 
mIR peak, which progressively shifts to lower energy 
with $P$, from 0.36~eV at $P$=2~GPa to 0.18~eV at 
$P$=20~GPa. 
As a check for the KK analysis, 
the $\sigma(0)$ values at 2 and 20~GPa, 
taken from the $\rho(T)$ study \cite{miyake}, 
are also displayed by the two dots in Fig.~1(c).  
As shown by the broken curves, the obtained $\sigma(\omega)$ 
connect reasonably with $\sigma(0)$, indicating the 
appropriateness of the KK analysis despite the 
lack of $R_d(\omega)$ data in the far IR range.  
As mentioned in Introduction, many previous 
studies on Yb-based HF compounds found similar 
mIR peaks.  However, the 
present result is remarkable in that the mIR 
peak is so well-separated from the Drude (free 
carrier) component peaked at zero energy.  
[Note that a Drude component is not seen in 
Fig.~1(c) since it is located at lower photon 
energies below our measurement range, as suggested 
by the broken curves.]  
As already mentioned, it has been found that the 
peak energy $E_{\rm mIR}$ is roughly proportional 
to the $c$-$f$ hybridization energy ($\widetilde{V}$) 
over many compounds with different values of 
$\widetilde{V}$ \cite{okamura}.    The hybridization is 
expected to be strongly $P$ dependent, so it is 
interesting to analyze the $P$ evolution of 
$E_{\rm mIR}$ in terms of $\widetilde{V}$.   
Although $\gamma$ data was used to estimate 
$\widetilde{V}$ in the previous study \cite{okamura}, 
$\gamma$ of YbCu$_2$Ge$_2$ at high $P$ is unavailable.  
However, $A$ data of YbCu$_2$Ge$_2$ at high $P$ 
are available \cite{miyake}.   
Since $A^{1/2}$ is proportional to $m^\ast$, it is 
inversely proportional to the Kondo temperature 
$T_{\rm K}$.   
The hybridization is expressed as 
$2\widetilde{V} \simeq \sqrt{W T_{\rm K}}$ \cite{cox,coleman}, 
where $W$ is the bandwidth of the $c$ band.  
Therefore, if $W$ does not change with $P$, we have 
$\widetilde{V} \propto A^{-1/4}$, and we may utilize 
the $A^{1/2}$ data to analyze the evolution of 
$E_{\rm mIR}$ in terms of $\widetilde{V}$.

Figure~2(a) compares the measured $P$ evolution of 
$E_{\rm mIR}$ with that of the reported $A^{1/2}$ data 
\cite{miyake}.  
\begin{figure}[t]
\includegraphics[width=0.8\textwidth]{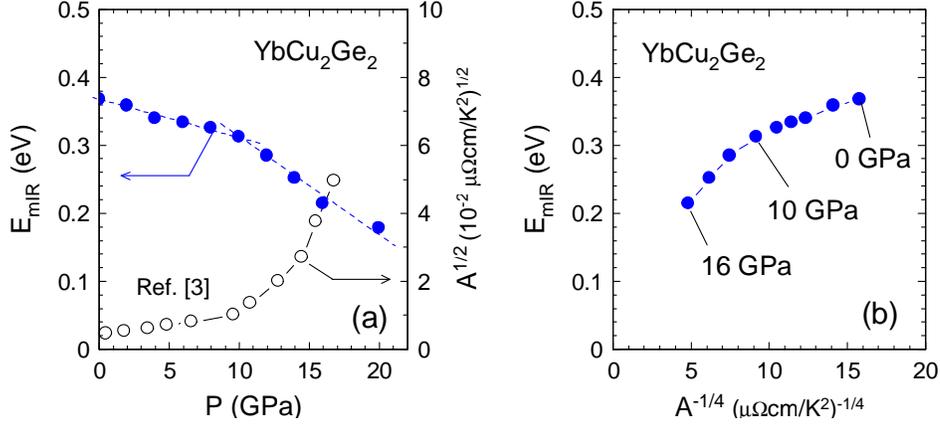}
\caption{(a)Measured peak energy ($E_{\rm mIR}$) of 
the mIR peak in $\sigma(\omega)$ and the $A^{1/2}$ 
data taken from Miyake {\it et al.} \cite{miyake} 
plotted as a function of pressure ($P$).  
The broken lines are guide to the eye, emphasizing the 
different $P$ dependences of $E_{\rm mIR}$ below and 
above 10~GPa.  
(b) $E_{\rm mIR}$ plotted as a function of $A^{-1/4}$.  
Here, $A^{-1/4}$ is proportional to $\sqrt{T_{\rm K}}$, 
which in turn is proportional to the renormalized 
hybridization energy $\widetilde{V}$.  
Hence this graph effectively compares $E_{\rm mIR}$ with 
the hybridization $\widetilde{V}$.   
}
\end{figure}
[$\sigma(\omega)$ at $P$=0 was measured without DAC, 
but not shown here.]  
Note that the decrease of $E_{\rm mIR}$ with $P$ is 
not monotonic, but that above 10~GPa is more rapid 
than that below 10~GPa as emphasized by the broken 
lines in Fig.~2(a).   A similar tendency is also seen 
for the $A^{1/2}$ data, namely the increase of $A^{1/2}$ 
above 10~GPa is more rapid than that below 10~GPa, 
indicating that $m^\ast$ increases with $P$ more rapidly 
above 10~GPa.   
Namely, both $E_{\rm mIR}$ and $m^\ast$ vary with $P$ more 
rapidly above 10~GPa.  Although $\sigma(\omega)$ 
was measured at 295~K and $A^{1/2}$ at much lower $T$, 
it seems likely that these results of $E_{\rm mIR}$ 
and $A^{1/2}$ are related to each other.  
Figure~2(b) shows plots of $E_{\rm mIR}$ as a function 
of $A^{-1/4}$.   As discussed above, 
$\widetilde{V} \propto A^{-1/4}$ if $W$ is 
independent of $P$.  Hence, Fig.~2(b) is effectively 
a graph of $E_{\rm mIR}$ as a function of $\widetilde{V}$.   
In the $c$-$f$ hybridized band model, 
$E_{\rm mIR} \simeq 2\widetilde{V}$, and if this model 
is applied here, $E_{\rm mIR} \propto A^{-1/4}$ is 
expected.    However, this is apparently not the case 
in Fig.~2(b), where $E_{\rm mIR}$ is not 
quite proportional to $A^{-1/4}$.  
Nevertheless, $E_{\rm mIR}$ is indeed an increasing 
function of $A^{-1/4}$, and the move of the plot 
to lower left with $P$ is consistent with a decrease 
of $\widetilde{V}$ expected from the observed 
increase of $A^{1/2}$ at high $P$.    
Furthermore, the dependence of $E_{\rm mIR}$ on $A^{-1/4}$ 
seems different between below and above 10~GPa.  Namely, 
$E_{\rm mIR}$ is almost linear in $A^{-1/4}$ with 
different slopes between below and above 10~GPa.  
Since the average Yb valence in YbCu$_2$Ge$_2$ at low $P$ 
is close to 2 (divalent), the hybridized band model may 
not be appropriate, which could be responsible for the 
deviation of $E_{\rm mIR}$ vs $A^{-1/4}$ graph from 
proportionality.  
In the photoemission spectrum of YbCu$_2$Ge$_2$ 
at $P$=0 \cite{PES}, a strong peak due to $f^{14}$ (Yb$^{2+}$) 
state has been observed about 0.2~eV below $E_{\rm F}$.   
Therefore, it is probably more natural to interpret the 
mIR peak in $\sigma(\omega)$ at low $P$ range as arising from 
excitations from this $f^{14}$ state to the unoccupied 
states above $E_{\rm F}$.  
Then, the lower-energy shifts of mIR peak with $P$ 
show that the $f^{14}$ level is moving up toward $E_{\rm F}$ 
with increasing $P$.  Such an upward shifts of $f^{14}$ level 
is expected by simple theoretical consideration \cite{mito}, 
but has rarely been observed by a spectroscopic 
experiment.  
At higher $P$ range, on the other hand, the evolutions 
of $E_{\rm mIR}$ may be more appropriately interpreted 
using the hybridized band model, since the $f^{14}$ state 
should be actually crossing $E_{\rm F}$.  
Then, the change of slope from below to above 10~GPa in 
Figs.~2(a) and 2(b) might correspond to the low and high 
$P$ ranges in the above scenario.    
More detailed analyses including the $T$ 
dependences of $\sigma(\omega)$ will be presented in a 
future publication.

\section{Summary}
$\sigma(\omega)$ spectra of YbCu$_2$Ge$_2$ at 295~K 
have been measured at high $P$ to 20~GPa.  
A pronounced mIR peak has been observed in $\sigma(\omega)$, 
and has shown significant shifts to lower energy with 
increasing $P$.  
The $P$ induced shifts of mIR peak has been analyzed 
in terms of the $c$-$f$ hybridization energy, using 
the $T^2$ coefficient of $\rho(T)$.  It is suggested 
that the $P$ evolution of $E_{\rm mIR}$ at low $P$ 
range corresponds 
to the upward shift of $f^{14}$ state toward $E_{\rm F}$ 
with $P$, and that at high $P$ range to the evolution 
of the $c$-$f$ hybridized bands.

\section{Acknowlegment}
Experiments at SPring-8 were performed under approval 
by JASRI (2014B1751, 2014B1749, 2015B1698, 2015B1697). 
H.~O. acknowledges financial support from JSPS KAKENHI 
(21102512, 23540409, 26400358).

\end{document}